\documentclass[final]{aipproc}

\def\gs{\mathrel{\raise0.35ex\hbox{$\scriptstyle >$}\kern-0.6em
\lower0.40ex\hbox{{$\scriptstyle \sim$}}}}
\def\ls{\mathrel{\raise0.35ex\hbox{$\scriptstyle <$}\kern-0.6em
\lower0.40ex\hbox{{$\scriptstyle \sim$}}}}

\layoutstyle{6x9}

\SetInternalRegister\hbadness{8000} 

\newcommand\doingARLO[2][]{
  \ifx\mmref\undefined #1\else #2\fi
}
  
\def    \beq       {\begin{equation}}
\def    \eeq       {\end{equation}}

\def    \cm     {\,{\rm cm}}

\def    \erg    {\,{\rm erg}}
\def	\g	{\,{\rm g}}

\def    \K      {\,{\rm K}}

\def    \ltsim  {\lesssim}      
\def    \gtsim  {\gtrsim}       

\def    \s      {\,{\rm s}}

\def    \sr     {\,{\rm sr}}

\def	\mum	{\,\mu{\rm m}}

\def	\simali	{\sim\,}

\def    \Qext   {Q_{\rm ext}}
\def    \Qabs   {Q_{\rm abs}}

\def    \cext   {C_{\rm ext}}
\def    \cabs   {C_{\rm abs}}

\def    \cgeo   {C_{\rm geo}}

\def    \ltsim  {\lesssim}      
\def    \gtsim  {\gtrsim}       
\def    \simlt  {\lesssim}      
\def    \kabs   {\kappa_{\rm abs}}
\def    \emt    {\epsilon_{\lambda}}
\def    \qabsa  {\left(Q_{\rm abs}/a\right)}
\def    \md     {m_{\rm dust}}
\def    \Td     {T_{\rm d}}
\newcommand	  \epsre	{\epsilon_1}
\newcommand	  \epsim	{\epsilon_2}

\def\lesssim{\mathrel{\hbox{\rlap{\hbox{\lower4pt\hbox{$\sim$}}}\hbox{$<$}}}}
\def\gtrsim{\mathrel{\hbox{\rlap{\hbox{\lower4pt\hbox{$\sim$}}}\hbox{$>$}}}}

\begin{document}

\title 
      [Absorption and Emission Properties of Interstellar Grains]
      {On the Absorption and Emission Properties of Interstellar Grains}


\author{Aigen Li}{
  address={Department of Physics and Astronomy, 
  University of Missouri, Columbia, MO 65211, USA\\
  email: {\sf LiA@missouri.edu}}
}

\vspace*{-3.6em}
{\noindent \normalsize\tt invited talk for
{\bf ``The Spectral Energy Distribution of Gas-Rich 
Galaxies: Confronting Models with Data''} 
(Heidelberg, Germany/4--8 October 2004), edited by
C.C. Popescu \& R.J. Tuffs, AIP Conf. Ser., in press}
\vspace*{-1.7em}

\copyrightyear  {2005}

\begin{abstract}
Our current understanding of the absorption and emission
properties of interstellar grains are reviewed. 
The constraints placed by the Kramers-Kronig relation on 
the wavelength-dependence and the maximum allowable
quantity of the dust absorption are discussed.
Comparisons of the opacities (mass absorption coefficients)
derived from interstellar dust models with those directly 
estimated from observations are presented.  
\end{abstract}

\date{\today}

\maketitle

\section{1.~~ Introduction}
Interstellar dust reveals its presence in astrophysical
environments and its (both positive and negative) role in
astrophysics mainly through its interaction with electromagnetic 
radiation (see Li [38] for a recent review): 
\vspace{-2mm}
\begin{itemize}
\item {\it obscuring} distant stars by the {\it absorption} 
and {\it scattering} of starlight by dust (the combined effects 
of absorption and scattering are called {\it extinction}); 
\item {\it reddening} starlight because the extinction is 
stronger for blue light than for red; 
\item generating ``{\it  reflection nebulae}'' 
      by the {\it scattering} of starlight by dust 
      in interstellar clouds near one or more bright stars;
\item generating the ``{\it diffuse Galactic light''} 
      seen in all directions in the sky by the diffuse
      scattering of starlight of stars located near 
      the Galactic plane;
\item generating {\it X-ray halos} by the small-angle dust 
      scattering of X-ray sources;
\item {\it polarizing} starlight as a result of preferential 
      extinction of one linear polarization over another 
      by aligned nonspherical dust; 
\item {\it heating} the interstellar gas by ejecting photoelectrons
      created by the absorption of energetic photons; 
\item and {\it radiating} away the absorbed short-wavelength
      radiation at longer wavelengths from near infrared (IR)  
      to millimeter (mm) in the form of {\it thermal emission}, 
      with a small fraction at far-red wavelengths 
      as {\it luminescence}.
\end{itemize}

In order to correct for the effects of interstellar
extinction and deredden the reddened starlight,
it is essential to understand the absorption and scattering 
properties of interstellar grains at short wavelengths
(particularly in the optical and ultraviolet [UV]). 
The knowledge of the optical and UV properties 
of interstellar dust is also essential 
for interstellar chemistry modeling 
since the attenuation of UV photons 
by dust in molecular clouds protects molecules 
from being photodissociated. 
The knowledge of the dust emission properties at longer 
wavelengths are important (i) for interpreting the IR 
and submillimeter (submm) observations of emission from 
dust and tracing the physical conditions of the emitting
regions, (ii) for understanding the process of star formation 
for which the dust is not only a building block but also
radiates away the gravitational energy of collapsing 
clouds (in the form of IR emission) and therefore making 
star formation possible, and (iii) for understanding 
the heating and cooling of the interstellar medium (ISM) 
for which interstellar dust is a dominant heating source 
by providing photoelectrons (in the diffuse ISM)
and an important cooling agent in dense regions
by radiating in the IR (see Li \& Greenberg [45] for a review).

Ideally, if we know the size, shape, geometry
and chemical composition (and therefore the dielectric
function) of an interstellar grain, we can calculate
its absorption and scattering cross sections as a function 
of wavelength. If we also know the intensity of the illuminating
radiation field, we should be able to calculate the equilibrium
temperature or temperature distribution of the grain from
its absorption cross section and therefore predict its IR 
emission spectrum. 

However, our current knowledge of the grain size, 
shape, geometry and chemical composition
is very limited; the nature of interstellar dust
itself is actually mainly derived from its interaction 
with radiation (see Li [38] for a review):
\vspace{-2mm}
\begin{itemize}
\item from the interstellar extinction curve which displays 
an almost linear rise with inverse wavelength ($\lambda^{-1}$)
from the near-IR to the near-UV and a steep rise into
the far-UV one can conclude that interstellar grains must
span a wide range of sizes, containing appreciable numbers
of {\it submicron}-sized grains 
as well as {\it nanometer}-sized grains;
\item from the wavelength dependence of the interstellar 
polarization which peaks at $\lambda \sim 0.55\mum$,
one can conclude that some fraction of the interstellar grains 
must be nonspherical and aligned by some process, 
with a characteristic size of $\sim 0.1\mum$; 
\item from the scattering properties measured for
interstellar dust which are characterized by
a quite high albedo ($\simali$0.6) in the near-IR 
and optical and a quite high asymmetry factor
(typically $\sim$0.6--0.8 in the optical) one can
infer that a considerable fraction of the dust must
be dielectric and the predominantly forward-scattering
grains are in the submicron size range;
\item from the IR emission spectrum of the
diffuse ISM which is characterized by a modified black-body 
of $\lambda^{-1.7}\,B_\lambda$(T=19.5$\K$)
peaking at $\sim 130 \mum$ in the wavelength range
of 80$\mum \ltsim \lambda \ltsim 1000\mum$,
and a substantial amount of emission at 
$\lambda\simlt60\mum$ which far exceeds what 
would be expected from dust at $T \approx20\K$,
one can conclude that in the diffuse ISM, 
(1) the bulk interstellar dust is in 
the {\it submicron} size range and heated to 
an {\it equilibrium temperature} around 
$T\sim 20\K$, responsible for the emission 
at $\lambda\gtsim 60\mum$; and (2) there also exists 
an appreciable amount of {\it ultrasmall} grains in 
the size range of a few angstrom to a few nanometers 
which are {\it stochastically heated by single UV photons} 
to high temperatures ($T>50\K$), responsible for the 
emission at $\lambda\simlt 60\mum$ (see Li [39]);
\item from the spectroscopic absorption features at 
9.7, 18$\mum$ and 3.4$\mum$ and emission features
at 3.3, 6.2, 7.7, 8.6 and 11.3$\mum$ which are 
collectively known as the ``UIR'' (unidentified IR) bands, 
one can conclude that interstellar dust consists 
of appreciable amounts of amorphous silicates 
(of which the Si-O stretching mode
and the O-Si-O bending mode are respectively responsible 
for the 9.7 and 18$\mum$ features), aliphatic hydrocarbon 
dust (of which the C-H stretching mode is responsible for
the 3.4$\mum$ feature), and aromatic hydrocarbon molecules
(of which the C-H and C-C stretching and bending vibrational
modes are responsible for the 3.3, 6.2, 7.7, 8.6 and 11.3$\mum$
``UIR'' features), although the exact nature of the carriers 
of the 3.4$\mum$ feature and the ``UIR'' features remain 
unknown. 
\end{itemize}
The inferences from observations for interstellar dust
summarized above are quite general and model-independent. 
But these inferences are not sufficient to quantitatively
derive the absorption and emission properties of interstellar
grains. For a quantitative investigation, one needs to make
{\it prior} specific assumptions concerning the grain size, 
shape, geometry and chemical composition which are still not
well constrained by the currently available observational data.  
To this end, one needs to adopt a specific grain model
in which the physical characteristics of interstellar dust
are fully specified. While a wide variety of grains models 
have been proposed to explain the interstellar extinction, 
scattering, polarization, IR emission and elemental 
depletion, so far no single model can satisfy
{\it all} the observational constraints 
(see Li [39] and Dwek [22] for recent reviews).

In view of this, in this article I will first try to 
{\it place constraints on the absorption and emission 
properties of interstellar dust based on general 
physical arguments; these constraints are essentially 
model-independent}. 

In astrophysical literature, the most frequently used
quantities describing the dust absorption and emission 
properties are the {\it mass absorption coefficient} 
(also known as ``{\it opacity}'') $\kabs$ with a unit
of $\cm^2\g^{-1}$, and the {\it emissivity} $\emt$, 
defined as the energy emitted per unit wavelength 
per unit time per unit solid angle per unit mass, 
with a unit of $\erg\s^{-1}\sr^{-1}\cm^{-1}\g^{-1}$.  
The Kirchhoff's law relates $\emt$ to $\kabs$ 
through $\emt = \kabs(\lambda)\,B_\lambda(T)$ if the dust
is large enough to attain an equilibrium temperature 
$T$ when exposed to the radiation field, or 
$\emt = \kabs(\lambda)\,\int_{0}^{\infty} dT\,B_\lambda(T)dP/dT$
if the dust is so small that it is subject to single-photon
heating and experiences ``temperature spikes'',
where $B_\lambda$ is the Planck function,
$dP$ is the probability for the dust to have 
a temperature in $[T,\,T+dT]$.  
Other often used quantities are the 
{\it absorption cross section} $\cabs$
and {absorption efficiency} $\Qabs$,
with the latter defined as the absorption
cross section $\cabs$ divided by the geometrical 
cross sectional area $\cgeo$ of the grain projected
onto a plane perpendicular to the incident
electromagnetic radiation beam (Bohren \& Huffman [8]). 
For spherical grains of radii $a$, $\cgeo=\pi a^2$ 
so that $\Qabs=\cabs/\pi a^2$.
By definition, $\kabs=\cabs/m=\cabs/\left(V\rho\right)$,
where $m$, $V$ and $\rho$ are respectively
the dust mass, volume, and mass density;
for spherical grains, 
$\kabs=3\Qabs/\left(4a\rho\right)$.

In \S2 I will apply the Kramers-Kronig relation
to place a lower limit on $\beta$ (the wavelength 
dependence exponent index of $\kabs$) and an upper 
limit on the absolute value of $\kabs$.
The state of our knowledge of interstellar grain opacity 
will be presented in \S3 (with a focus on $\beta$) and 
in \S4 (with a focus on the absolute value of $\kabs$),
followed by a summary in \S5.

\vspace{-1em}
\section{2.~~ Constraints from the Kramers-Kronig Relation}
As already mentioned in \S1, with specific assumptions
made concerning the grain size, shape, geometry and 
composition, {\it in principle} one can calculate 
the absorption cross section $\cabs$ and the opacity $\kabs$
as a function of wavelength. But this is limited
to spherical grains; even for grains with such simple 
shapes as spheroids and cylinders, the calculation is  
complicated and limited to small size parameters
defined as $2\pi a/\lambda$. As a result, astronomers 
often adopt a simplified formula
\begin{equation}\label{eq:kabs}
\kabs(\lambda) = 
\left\{\begin{array}{lr}
\kappa_0~, &\lambda < \lambda_0~,\\
\kappa_0\,\left(\lambda/\lambda_0\right)^{-\beta}~, 
&\lambda \ge \lambda_0~,\\
\end{array}\right.
\end{equation}
where $\lambda_0$ and the exponent index
$\beta$ are usually treated as free parameters,
while $\kappa_0$ is often taken from experimental
measurements of cosmic dust analogs or values 
predicted from interstellar dust models; or
\begin{equation}\label{eq:qabs}
\qabsa(\lambda) = 
\left\{\begin{array}{lr}
\qabsa_0~, &\lambda < \lambda_0~,\\
\qabsa_0\,\left(\lambda/\lambda_0\right)^{-\beta}~, 
&\lambda \ge \lambda_0~,\\
\end{array}\right.
\end{equation}
where again $\lambda_0$ and $\beta$ are free 
parameters and $\qabsa_0$ is usually taken from
model calculations.

The Kramers-Kronig dispersion relation, originally
deduced by Kronig [33] and Kramers [32] from
the classical Lorentz theory of dispersion of light,
connects the real ($m^{\prime}$; dispersive) and 
imaginary ($m^{\prime\prime}$; absorptive)
parts of the index of refraction 
($m[\lambda]=m^{\prime}+i\,m^{\prime\prime}$) based on 
the fundamental requirement of {\it causality}.
Purcell [55] found that the Kramers-Kronig relation 
can be used to relate the extinction cross section 
integrated over the entire wavelength range 
to the dust volume $V$
\begin{equation}\label{eq:kk1}
\int_{0}^{\infty} \cext(\lambda)\,d\lambda 
= 3 \pi^2 F V~~~,
\end{equation}
where $\cext$ is the extinction cross section, and $F$,  
a dimensionless factor, is the orientationally-averaged 
polarizability relative to the polarizability of an equal-volume 
conducting sphere, depending only upon the grain shape and 
the static (zero-frequency) dielectric constant 
$\epsilon_0$ of the grain material 
(Purcell [55]; Draine [18]).
Since $\cext$ is the sum of the absorption $\cabs$
and scattering $\cabs$ cross sections both of which
are positive numbers at all wavelengths, replacing
$\cext$ by $\cabs$ in the left-hand-side of Eq.(\ref{eq:kk1})
would give a lower limit on its right-hand-side;
therefore we can write Eq.(\ref{eq:kk1}) as
\begin{equation}\label{eq:kk2}
\int_{0}^{\infty} \kabs(\lambda)\,d\lambda 
< 3 \pi^2 F/\rho~~~.
\end{equation}
It is immediately seen in Eq.(\ref{eq:kk2}) 
that $\beta$ should be larger than 1 for 
$\lambda \rightarrow \infty$ since $F$ is a finite number 
and the integration in the left-hand-side of Eq.(\ref{eq:kk2}) 
should be convergent (also see Emerson [23]), 
although we cannot rule out $\beta\le 1$ 
over a finite range of wavelengths. 

Astronomers often use the opacity $\kabs$ of the formula described
in Eq.(\ref{eq:kabs}) to fit the far-IR, submm and mm photometric
data and then estimate the dust mass of interstellar clouds:
\begin{equation}\label{eq:md}
\md = \frac{d^2 F_\lambda}{\kabs(\lambda)\,B_\lambda(T)} ~~,
\end{equation}
where $d$ is the distance of the cloud, $T$ is the dust
temperature, $F_\lambda$ is the measured flux density at 
wavelength $\lambda$. By fitting the photometric data points, 
one first derives the best-fit parameters $\beta$, $\lambda_0$, 
and $T$. For a given $\kappa_0$, one then estimates $\md$ from 
Eq.(\ref{eq:md}). In this way, various groups of astronomers 
have reported the detection of appreciable amounts of very 
cold dust ($T<10\K$) both in the Milky Way and in external galaxies
(Reach et al.\ [56]; Chini et al.\ [14]; Kr\"ugel et al.\ [35];
Siebenmorgen, Kr\"ugel, \& Chini [59];
Boulanger et al.\ [10]; Popescu et al.\ [54]; Galliano et al.\ [24];
Dumke, Krause, \& Wielebinski [20]). 

As can be seen in Eq.(\ref{eq:md}), if the dust temperature
is very low, one then has to invoke a large amount of dust
to account for the measured flux densities. 
This often leads to too large a dust-to-gas ratio
to be consistent with that expected from the metallicity
of the region where the very cold dust is detected
(e.g. see Dumke et al.\ [20]), 
unless the opacity $\kabs$ is very large.
It has been suggested that such a large $\kabs$
can be achieved by fractal or porous dust 
(Reach et al.\ [56]; Dumke et al.\ [20]).
Can $\kabs$ be really so large for physically 
realistic grains? 
At a first glance of Eq.(\ref{eq:kk2}),
this appears plausible if the dust is sufficiently porous 
(so that its mass density $\rho$ is sufficiently small). 
However, one should keep in mind that for a porous grain, 
the decrease in $\rho$ will be offset by a decrease 
in $F$ because the effective static dielectric constant 
$\epsilon_0$ becomes smaller when the dust becomes porous, 
leading to a smaller $F$ factor 
(see Fig.\,1 in Purcell [55] and Fig.\,15 in Draine [18]).

Similarly, if one would rather use $Q_{\rm abs}/a$ of 
Eq.(\ref{eq:qabs}) instead of $\kabs$ of Eq.(\ref{eq:kabs}),
we can also apply the Kramers-Kronig relation to place 
(1) a lower limit on $\beta$ -- $\beta$ cannot be smaller than 
or equal to 1 at all wavelengths,
and (2) an upper limit on $\qabsa_0$ from
\begin{equation}\label{eq:kk3}
\int_0^{\infty} \qabsa(\lambda)\,d\lambda < 4\pi^2 F ~~.
\end{equation}
Finally, the best-fit parameters $\beta$, $\lambda_0$,
and $T$ should be physically reasonable. This can be 
checked by comparing the best-fit temperature $T$ with
the dust equilibrium temperature $\Td$ calculated from
the energy balance between absorption and emission
\begin{equation}\label{eq:calcT}
\int^{\infty}_{0} \kabs(\lambda)\, c u_{\lambda}\,d\lambda
= \int^{\infty}_{0} \kabs(\lambda) 
  4\pi B_{\lambda}(\Td)\,d\lambda~~,
\end{equation}
where $c$ is the speed of light,
and $u_{\lambda}$ is the energy density of the radiation field.
Alternatively, one can check whether the strength of the radiation 
field required by $\Td\approx T$ is in good agreement with 
the physical conditions of the environment where the dust
is located. 

\section{3. Opacity: Wavelength-dependence Exponent Index}
It is seen in \S2 that the Kramers-Kronig relation
requires $\beta >1$ for $\lambda \rightarrow \infty$.
For the Milky Way diffuse ISM, the 100$\mum$--1\,mm dust 
emission spectrum obtained by
the {\it Diffuse Infrared Background Experiment} (DIRBE) 
instrument on the {\it Cosmic Background Explorer} (COBE) 
satellite is well fitted by the product of a single Planck 
curve of $T\approx 17.5\K$ and an opacity law characterized 
by $\beta\approx 2$ (i.e. $\kabs \propto \lambda^{-2}$; 
Boulanger et al.\ [10]), although other sets of $T$ 
and $\beta$ are also able to provide (almost) equally 
good fits to the observed emission spectrum 
(e.g. $T\approx 19.5\K$ and $\beta\approx 1.7$; see Draine [17]). 
But we should emphasize here that it is never flatter 
than $\beta\approx 1.65$ (Draine [17]).

Smaller $\beta$ in the submm and mm wavelength range
has been reported for cold molecular cores 
(e.g. Walker et al.\ [63]: $\beta \approx 0.9-1.8$),
circumstellar disks around young stars
(e.g. Beckwith \& Sargent [4]: $\beta \approx -1$--1; 
Mannings \& Emerson [47]: $\beta \approx 0.6$;
Koerner, Chandler, \& Sargent [28]: $\beta\approx 0.6$),
and circumstellar envelopes around evolved stars
(e.g. Knapp, Sandell, \& Robson [27]: $\beta\approx 0.9$)
including the prototypical carbon star IRC\,+\,10126
(Campbell et al.\ [13]: $\beta\approx 1$).

However, these results are not unique since the dust 
temperature and density gradients in the clouds, disks
or envelopes have not been taken into account 
in deriving $\beta$ -- the $\beta$ exponent was 
usually estimated by fitting the submm and mm spectral 
energy distribution by a modified black-body 
$\lambda^{-\beta}\,B_\lambda(T)$ under the ``optically-thin''
assumption, with $\beta$ and $T$ as adjustable parameters. 
If the dust spatial distribution is not constrained,
the very same emission spectrum can be equally well fitted
by models with different $\beta$ values. 

As a matter of fact, an asymptotic value of $\beta\approx 2$ 
(i.e. $\kabs \propto \lambda^{-2}$) is expected for both 
dielectric and conducting {\it spherical} grains: 
in the Rayleigh regime (where the wavelength is much 
larger than the grain size)
\begin{equation}
\kabs \approx \frac{18\pi}{\lambda\rho}
\frac{\epsim}{(\epsre+2)^2+\epsim^2} ~~~,
\end{equation}
where $\epsilon(\lambda) = \epsre + i\,\epsim$ is the complex 
dielectric function of the grain at wavelength $\lambda$. 
For dielectric spheres, $\kabs \propto \lambda^{-1}\epsim \propto 
\lambda^{-2}$ as $\lambda\rightarrow \infty$ 
since $\epsre$ approaches a constant ($\gg\epsim$) 
while $\epsim \propto \lambda^{-1}$;
for metallic spheres with a conductivity of $\sigma$, 
$\kabs\propto \lambda^{-1}\epsim^{-1}\propto \lambda^{-2}$ 
as $\lambda\rightarrow \infty$ 
since $\epsim = 2\lambda\sigma/c \propto \lambda$ 
and $\epsre \ll \epsim$.
Even for {\it dielectric} grains with such an extreme
shape as needle-like prolate spheroids
(of semiaxes $l$ along the symmetry axis
and $a$ perpendicular to the symmetry axis),
in the Rayleigh regime we expect 
$\kabs \propto \lambda^{-2}$:
\begin{equation}\label{eq:needle}
\kabs \approx \frac{2\pi}{3\lambda\rho}\frac{\epsim}
{\left[L_{\|}(\epsre-1) + 1\right]^2 
+ \left(L_{\|}\epsim\right)^2}
\end{equation} 
where $L_{\|}\approx \left(a/l\right)^2\ln(l/a)$ is the
depolarization factor parallels to the symmetry axis;
since for dielectric needles $\epsre \rightarrow$ constant
and $L_{\|}(\epsre-1) + 1 \gg L_{\|}\epsim$,
therefore $\kabs \propto \lambda^{-1}\epsim
\propto \lambda^{-2}$ (see Li [36] for a detailed
discussion). Only for {\it both} conducting {\it and}
extremely-shaped grains $\kabs$ can still be large at
very long wavelengths. But even for those grains, 
the Kramers-Kronig relation places an upper limit 
on the wavelength range over which large $\kabs$
can be attainable (see Li \& Dwek [43] for details).
The Kramers-Kronig relation has also been applied to
interstellar dust models to see if the subsolar interstellar
abundance problem can be solved by fluffy dust (Li [40])
and to TiC nanoparticles to relate their UV absorption
strength to their quantities in protoplanetary nebulae (Li [37]).

The inverse-square dependence of $\kabs$ on wavelength 
derived above applies to both crystalline and amorphous
materials (see Tielens \& Allamandola [61] for a detailed
discussion). Exceptions to this are amorphous layered materials  
and very small amorphous grains in both of which the phonons are
limited to two dimensions and their phonon spectrum is
thus proportional to the frequency. Therefore, for both 
amorphous layered materials and very small amorphous grains 
the far-IR opacity is in inverse proportional to
wavelength, i.e. $\kabs \propto \lambda^{-1}$
(Seki \& Yamamoto [58]; Tielens \& Allamandola [61]).
Indeed, the experimentally measured far-IR absorption 
spectrum of amorphous carbon shows 
a $\kabs \propto \lambda^{-1}$ dependence at 
$5\mum < \lambda <340\mum$ (Koike, Hasegawa, \& Manabe [29]).
If there is some degree of cross-linking between the layers
in the amorphous layered grains, we would expect 
$1<\beta<2$ (Tielens \& Allamandola [61]). 
This can explain the experimental far-IR absorption spectra 
of layer-lattice silicates which were found to have 
$1.25<\beta <1.5$ at $50\mum <\lambda <300\mum$ (Day [15]). 
For very small amorphous grains, if the IR absorption
due to internal bulk modes (for which the density of
states frequency spectrum is proportional to $\lambda^{-2}$)
is not negligible compared to that due to surface
vibrational modes (for which the frequency spectrum is 
proportional to $\lambda^{-1}$),
we would also expect $1<\beta<2$ (Seki \& Yamamoto [57]).

If there exists a distribution of grain sizes,
ranging from small grains in the Rayleigh regime
for which $\beta\sim 2$ and very large grains
in the geometric optics limit for which $\beta \sim 0$,
we would expect $\beta$ to be intermediate between
0 and 2. This can explain the small $\beta$ values
of dense regions such as molecular cloud cores, 
protostellar nebulae and protoplanetary disks 
where grain growth occurs (e.g. see Miyake \& Nakagawa [50]).

The exponent index $\beta$ is temperature-dependent, 
as measured by Agladze et al.\ [1] for silicates
at $T=1.2$--30$\K$ and $\lambda=700$--2900$\mum$,
and by Mennella et al.\ [49] for silicates and 
carbon dust at $T=24$--295$\K$ and $\lambda=20$--2000$\mum$.
Agladze et al.\ [1] found that at $T=1.2$--30$\K$,
$\beta$ first increases with increasing $T$, after reaching 
a maximum at $T\sim 10$--15$\K$ it starts to decrease 
with increasing $T$. Agladze et al.\ [1] attributed 
this to a two-level population effect (B\"osch [9]):
because of the temperature-dependence of the two-level
density of states (i.e. the variation in temperature results
in the population change between the two levels), 
the exponent index $\beta$ is also temperature dependent.
In contrast, Mennella et al.\ [49] found that $\beta$ 
increases by $\sim$10\%--50\% from $T=295\K$ to 24$\K$, 
depending on the grain material (e.g. the variation of $\beta$ 
with $T$ for crystalline silicates is not as marked as for
amorphous silicates). The increase of $\beta$ with decreasing
$T$ at $T=24$--295$\K$ is due to the weakening of the long
wavelength absorption as $T$ decreases because at lower
temperatures fewer vibrational modes are activated. 
Finally, it is noteworthy that the inverse temperature
dependence of $\beta$ has been reported by Dupac et al.\ [21]
for a variety of regions.

\section{4.~~ Opacity: Absolute Values}
In literature, one of the widely adopted opacities is that
of Hildebrand [26]:
\begin{equation}\label{eq:kabs_H83}
\kabs\,({\rm cm}^2\g^{-1}) \approx 
\left\{\begin{array}{lr} 
2.50\times 10^3\left(\lambda/\mu {\rm m}\right)^{-1},
& 50\mum < \lambda \le 250\mum ~~,\\
6.25\times 10^5\left(\lambda/\mu {\rm m}\right)^{-2},
& \lambda > 250\mum ~~.\\
\end{array}\right.
\end{equation}
Hildebrand [26] arrived at the above values from
first estimating $\kabs(125\mu {\rm m})$ 
and then assuming $\beta\approx 1$ for $\lambda <250\mum$ 
and $\beta\approx 2$ for $\lambda >250\mum$.
He estimated the 125$\mum$ opacity from
$\kabs(125\mu {\rm m})= 3\Qabs(125\mu {\rm m})/\left(4a\rho\right)
=3/\left(4a\rho\right)
\left[\Qabs(125\mu {\rm m})/\Qext({\rm UV})\right] 
\Qext({\rm UV})$ by taking $a=0.1\mum$, $\rho=3\g\cm^{-3}$, 
$\Qext({\rm UV})=3$, 
and $\Qext({\rm UV})/\Qabs(125\mu {\rm m}) = 4000$,
where $\Qext({\rm UV})$ is the ultraviolet extinction
efficiency at $\lambda\simali$0.15--0.30$\mum$.

The most recent silicate-graphite-PAHs interstellar dust
model for the diffuse ISM (Li \& Draine [41]) gives 
\begin{equation}\label{eq:kabs_LD01}
\kabs\,({\rm cm}^2\g^{-1}) \approx 
\left\{\begin{array}{lr} 
2.92\times 10^5\left(\lambda/\mu {\rm m}\right)^{-2},
& 20\mum < \lambda \le 700\mum ~~,\\
3.58\times 10^4\left(\lambda/\mu {\rm m}\right)^{-1.68},
& 700\mum < \lambda \le 10^4\mum ~~,\\
\end{array}\right.
\end{equation}
while $\kabs(\lambda)\approx 4.6\times 10^{5}
\left(\lambda/\mu {\rm m}\right)^{-2}\cm^2\g^{-1}$
for the classical Draine \& Lee [19] silicate-graphite model.
The fluffy composite dust model of Mathis \& Whiffen [48]
has $\kabs(\lambda)\approx 2.4\times 10^{5}
\left(\lambda/\mu {\rm m}\right)^{-1.6}\cm^2\g^{-1}$
in the wavelength range of $100\mum < \lambda < 1000\mum$.
The silicate core-carbonaceous mantle dust model of
Li \& Greenberg [44] gives $\kabs(\lambda)\approx 1.8\times 10^{5}
\left(\lambda/\mu {\rm m}\right)^{-2}\cm^2\g^{-1}$
for $30\mum < \lambda < 1000\mum$. We see that
these $\kabs$ values differ by over one order-of-magnitude;
e.g., the 1350$\mum$ opacity calculated from the composite
model [$\kabs(1350\mu {\rm m})\approx 2.35\cm^2\g^{-1}$;
Mathis \& Whiffen [48]] is higher than that from 
the silicate-graphite-PAHs model 
[$\kabs(1350\mu {\rm m})\approx 0.20\cm^2\g^{-1}$;
Li \& Draine [41]] by a factor of $\sim$12.

While the Mathis \& Whiffen [48] composite dust model
predicts an IR emission spectrum too flat to be consistent
with the COBE-FIRAS observational spectrum, 
and the dust IR emission has not been calculated for
the Li \& Greenberg [44] core-mantle model
which focuses on the near-IR to far-UV extinction
and polarization, the silicate-graphite-PAHs model has 
been shown successful in reproducing the infrared 
emission spectra observed for the Milky Way
(Li \& Draine [41]), the Small Magellanic Cloud
(Li \& Draine [42]), and the ringed Sb galaxy NGC\,7331
(Regan et al.\ [57], Smith et al.\ [60]). 
Therefore, at this moment the dust opacity calculated 
from the silicate-graphite-PAHs model is preferred. 

It has recently been suggested that the long wavelength
opacity can be estimated from the comparison of the visual 
or near-IR optical depth with the (optically thin) far-IR, 
submm and mm dust emission measured for the same region with 
high angular resolution, assuming that both the short wavelength
extinction and the long wavelength emission are caused
by the same dust (e.g. see Alton et al.\ [2,3],
Bianchi et al.\ [5,6], Cambr\'esy et al.\ [12],  
Kramer et al.\ [30,31]) 
\begin{equation}\label{eq:alton}
\frac{\tau_V}{S(\lambda)} 
= \frac{\Qext(V)}{\Qabs(\lambda)}
  \frac{2.2\times 10^{-18}}{B_\lambda(T)}
\end{equation}
where $\tau_V$ is the visual optical depth, 
$S(\lambda)$ is the surface brightness at wavelength $\lambda$,
and $\Qext(V)$ is the extinction efficiency in
the $V$-band ($\lambda = 5500$\,\AA).
With the dust temperature $T$ determined from 
a modified black-body $\lambda^{-\beta}\,B_\lambda(T)$
fit to the far-IR dust emission spectrum 
($\beta$ is not treated as a free parameter but
taken to be a chosen number), $\Qabs(\lambda)$ can be 
calculated from the far-IR, submm, or mm surface density 
$S(\lambda)$, and the measured visual optical depth 
$\tau_V$ [if what is measured is the near-IR color-excess, 
say $E(H-K)$, instead of $\tau_V$, one can derive $\tau_V$ 
from $\tau_V \approx 14.6 E(H-K)$]. 
In so doing, $\Qext(V)$ is usually taken to be $\approx 1.5$.

The long wavelength $\kabs$ values recently determined
using this method (Eq.[\ref{eq:alton}]) are generally higher 
than those predicted from the dust models for the diffuse ISM.  
Although this can be explained by the fact that we are 
probably looking at dust in dense regions where the dust
has accreted an ice mantle and coagulated into fluffy 
aggregates for which a higher $\kabs$ is expected
(e.g. see Kr\"ugel \& Siebenmorgen [34], 
Pollack et al.\ [52], Ossenkopf \& Henning [51], 
Henning \& Stognienko [25], Li \& Lunine [46]),
the method itself is subject to large uncertainties:
(1) the grains responsible for the visual/near-IR 
extinction may not be the same as those responsible for
the far-IR, submm and mm emission; the latter is more 
sensitive to large grains while the former is dominated 
by submicron-sized grains; (2) the dust temperature $T$ 
may have been underestimated if the actual $\beta$ is larger 
than chosen; and (3) the fact that in many cases the IRAS 
({\it Infrared Astronomical Satellite}) 60$\mum$ photometry 
was included in deriving the dust temperature $T$ results
in appreciable uncertainties since the 60$\mum$ emission is
dominated by stochastically heated ultrasmall grains;
ignoring the temperature distributions of those grains
would cause serious errors in estimating the dust mass
(see Draine [16]) and therefore also in deriving the long 
wavelength opacity $\kabs$. These problems could be solved
by a detailed radiative transfer treatment
of the interaction of the dust with starlight
(e.g. Popescu et al.\ [53], Tuffs et al.\ [62])
together with a physical interstellar dust model
(e.g. the silicate-graphite-PAHs model; 
see Li \& Draine [41,42]).

Based on the laboratory measurements of the far-IR
and mm absorption spectra of both amorphous and crystalline
silicates as well as disordered carbon dust as a function
of temperature, Agladze et al.\ [1] and Mennella et al.\ [49]
found that not only the wavelength dependence exponent index
$\beta$ but also the absolute values of the absorption 
are temperature dependent: the far-IR and mm opacity 
systematically decreases (almost linearly) with decreasing 
temperature to $T\sim$10--15$\K$ and then increases with 
decreasing temperature at very low temperature.  
While the linear dependence of $\kabs$ on $T$ at $T>10$--15$\K$
was interpreted by Mennella et al.\ [49] in terms of 
two-phonon difference processes, the inverse-temperature 
dependence of $\kabs$ on $T$ at very low temperature
was attributed to a two-level population effect
(Agladze et al.\ [1]).
Agladze et al.\ [1] and Mennella et al.\ [49]
also found that the far-IR and mm opacity of amorphous
materials are larger than that of their crystalline
counterparts. This is because for amorphous materials,
the loss of long-range order of the atomic arrangement
leads to a relaxation of the selection rules that govern 
the excitation of vibrational modes so that all modes are
infrared active, while for crystalline solids, only a small 
number of lattice vibrations are active.

\section{5.~~ Summary}
The wavelength-dependent mass absorption coefficient
(opacity $\kabs$) is a critical parameter in the determination 
of the total dust mass of IR-emitting dusty regions.
The dust opacity shows marked variation with local conditions.
Due to the incomplete understanding of the size, shape,
composition and structure of dust grains, our knowledge
of the long wavelength dust opacity is subject to large 
uncertainties. We apply the Kramers-Kronig relation to
place a lower limit on the exponent index $\beta$ and
an upper limit on the absolute value of the opacity
(\S2), if the dust opacity is described as a power-law 
function of wavelength ($\kabs \sim \lambda^{-\beta}$). 
Our current knowledge of the wavelength dependence 
exponent index $\beta$ (\S3) and the absolute values 
of the opacity (\S4) of interstellar dust is summarized
in the context of interstellar grain models,
laboratory measurements, and direct comparison of
the short-wavelength extinction with the long wavelength 
thermal emission.

\vspace{2mm}
\noindent{\small {\bf ACKNOWLEDGMENTS.} I am grateful to 
C.C. Popescu and R.J. Tuffs for inviting me to give an
invited talk at this stimulating conference from which 
I learned much. I also thank E. Dwek, C.C. Popescu, 
R.J. Tuffs, A.N. Witt, and E.M. Xilouris for helpful 
discussions.}

\doingARLO[\bibliographystyle{aipproc}]
          {\ifthenelse{\equal{\AIPstyleselect}{num}}
             {\bibliographystyle{arlonum}}
             {\bibliographystyle{arlobib}}
          }


\end{document}